\documentclass{article}
\begin{document}
\title{\large \bf LOCALIZATION OF FIELDS ON A BRANE IN SIX DIMENSIONS}
\author{{\bf Merab Gogberashvili$^a$ and Paul Midodashvili$^b$} \\
$^a$ Andronikashvili Institute of Physics, 6 Tamarashvili Str.,
Tbilisi 380077, Georgia \\
{\sl E-mail: gogber@hotmail.com }\\
$^b$ Tskhinvali State University, 1 Cherkezishvili Str.,
Gori 383500, Georgia \\
{\sl E-mail: midodashvili@hotmail.com } }
\maketitle
\begin{abstract}
\quotation{Universe is considered as a brane in infinite (2+4)-space. 
It is shown that zero modes of all kinds of matter fields and
4-gravity are localized on the brane by increasing transversal
gravitational potential.}
\end{abstract}
\vskip 0.5cm

The papers \cite{ADD,RaSu,G1} had excited recent interest in brane
models. In present paper we want to concentrate on the 
localization problem in the model were our world is considered as
a single shell expanding in multi-dimensions
\cite{G2,G3,G4,G5,GoMi}. Two observed facts of modern cosmology,
the isotropic runaway of galaxies and the existence of a preferred
frame in the Universe where the relict background radiation is
isotropic, have the obvious explanation in this picture.

We assume that trapping of physical fields on the brane has the
gravitational nature, since in our world gravity is known to be
the unique interaction which has universal coupling with all
matter fields. To provide universal and stable trapping we assume
also that on the brane (where all gravitating matter can be
resided) gravitational potential should have minimal value with
the respect of extra coordinates. Growing gravitational potential
(warp factor) is the opposite choice compared to the one of
Randall-Sundrum with the maximum on the brane \cite{RaSu}.
However, Newton's law on the brane is the result of the
cancellation mechanism introduced in \cite{G1,G4} which allows
both types of gravitational potential.

To have localized multi-dimensional fields on a brane "coupling"
constants appearing after integration of their Lagrangian over
extra coordinates must be non-vanishing and finite. In
(1+4)-dimensional models following facts were clarified: spin $0$
field is localized on the brane with decreasing warp factor and
spin $1/2$ field - on the brane with increasing warp factor
\cite{BaGa}; spin $1$ field is not normalizable at all \cite{Po}
and spin $2$ fields are localized on the brane with decreasing
warp factor \cite{RaSu,G1}. For the case of (1+5)-dimensions it
was found that spin $0$, $1$ and $2$ fields are localized on the
brane with decreasing warp factor and the spin $1/2$ field on the
brane with increasing warp factor \cite{Od}. So to fulfill the
localization of Standard Model particles in (1+4)-, or
(1+5)-spaces it is required to introduce other interaction but
gravity.

Here we want to show that zero modes of spin $0$, $1/2$, $1$ and $2$ 
fields can be all localized on the brane in the (2+4)-space by 
increasing warp factor. Our motivation for the choice of the 
signature of the bulk is as follows. In the massless field case 
(weakest coupling with gravity) symmetries of a multi-dimensional 
manifold can be restored. It is well known, that in the zero-mass 
limit the main equations of physics are invariant under the 
15-parameter nonlinear conformal transformations. A long time ago it 
was also discovered that the conformal group can be written as a 
linear Lorentz-type transformation in a (2+4)-space (for these 
subjects see, for example, \cite{PeRi}).

Action of the gravitating system in six dimensions can be written
in the form
\begin{equation} \label{action}
S = \int d^6x\sqrt{^6g}\left[- \frac{M^4}{2}(^6R + 2 \Lambda) +
^6L \right] ,
\end{equation}
where $^6g$ is the determinant, $M$ is the fundamental scale,
$^6R$ is the scalar curvature, $\Lambda$ is the cosmological
constant and $^6L$ is the Lagrangian of matter fields, all these
values refer to six dimensions. Einstein's 6-dimensional equations
can be written in the form
\begin{equation} \label{Einstein6}
^6R_{AB} = -\frac {1}{2}\Lambda g_{AB} + \frac {1}{M^4}\left(
T_{AB}- \frac {1}{4} g_{AB}T\right).
\end{equation}
Capital Latin indices run over $A, B,... = 0, 1, 2, 3, 5, 6 $.

It is convenient to introduce the new dimensionless coordinates
$z, v$ of extra (1+1)-space except of Cartesian ones
\begin{eqnarray} \label{coordinates}
x^5 = \epsilon\sqrt{z}\cosh{v} , ~~~~~ x^6 =
\epsilon\sqrt{z}\sinh{v} ,  \\
z =\frac{x^2_5-x^2_6}{\epsilon^2},~~~~\tanh{v} = \frac{x^6}{x^5} .
\nonumber
\end{eqnarray}
The constant $\epsilon $ which makes $z, v$ to be dimensionless
corresponds to the width of the brane.

We are looking for the solution of (\ref{Einstein6}) in the form
\begin{equation} \label{ansatz}
ds^2= \phi ^2(z) \eta_{\alpha \beta }(x^\nu)dx^\alpha dx^\beta +
g_{ij}(z)dx^i dx^j ,
\end{equation}
where Greek indices $\alpha, \beta,... = 0, 1, 2, 3$ numerate
coordinates in 4-dimensions, while small Latin indices
$i, j, ... = 5, 6$ - coordinates of the transversal space. It is
assumed that in ansatz (\ref{ansatz}) the 4-dimensional conformal
factor $\phi^2$ and the metric tensor of transversal (1+1)-space
$g_{ij}$ depend on the extra coordinates $x^i$ only via the
coordinate $z$.

Suppose also that the extra coordinates enter the stress-energy
$T_{AB}$ from the metric (\ref{ansatz}) only. This means that
strength of a gauge fields $A_B$ towards the extra directions
and covariant derivatives of scalar $\Phi$ and spinor $\Psi$ 
fields with respect to the extra coordinates are zero \cite{G3}
\begin{equation} \label{extra}
F_{iB} = 0 , ~~~~~D_i\Phi = 0 , ~~~~~D_i\Psi = 0.
\end{equation}
Then ansatz for multi-dimensional matter energy-momentum tensor can
be written in the form
\begin{equation} \label{source}
T_{\alpha\beta } = \frac{\tau_{\alpha\beta }(x^\nu)}{\epsilon^2
\phi ^2(z)}, ~~~~ T_{ij} = - g_{ij}(z)\frac{L(x^\nu)}{\epsilon^2
\phi ^4(z)} .
\end{equation}
Because of conformal mapping in the space (\ref{ansatz}) the
4-dimensional Lagrangian of matter fields $L(x^\nu)$ and the
4-dimensional stress-energy $\tau_{\alpha \beta}(x^\nu )$
automatically appear to be independent from $z$ (see for example
\cite{PeRi}).

So we are looking for the solution of 6-dimensional Einstein's and
matter fields equations for the case of brane-Universe when the
metric and matter energy-momentum tensor have the general structures
(\ref{ansatz}) and (\ref{source}) respectively. As it was shown in
\cite{G3} this configuration corresponds to the solution with the
minimal energy and thus is stable.

On the brane we require to have 4-dimensional Einstein's equations
without a cosmological term
\begin{equation} \label{Einstein4}
R_{\alpha\beta} = \frac{1}{\epsilon^2 M^4\phi^2 }\left( \tau
_{\alpha\beta} - \frac{1}{2}\eta_{\alpha\beta}\tau\right).
\end{equation}
The Ricci tensor in four dimensions $R_{\alpha\beta}$ is
constructed from the 4-dimensional metric tensor
$\eta_{\alpha\beta}(x^{\nu})$ in the standard way. Then remaining
in (\ref{Einstein6}) equations reduce to \cite{GoMi}
\begin{equation} \label{system}
g_{ij} = c\eta _{ij}\phi^{'} , ~~~~~ z\phi^3 \phi^{'} + A\phi^5 +
B\phi + C = 0 ,
\end{equation}
were prime denotes a derivative with respect to $z$. Here $\eta _{ij}$
is the metric tensor of flat extra (1+1)-space, $c$ and $C$ are
the integration constants and
\begin{equation} \label{parameters}
A = - \frac{\Lambda \epsilon^2 c}{40},~~~~~ B = \frac{c(\tau +
2L)}{16 M^4}
\end{equation}
are the dimensionless parameters. In general $B$ depends on the
4-coordinates $x^\nu$.

To localize matter on the brane without extra sources the factor
$1/\phi ^2(z)$ in (\ref{source}) and (\ref{Einstein4}) should have
$\delta$-like behavior. It means that $\phi^2(z)$ (and transversal
gravitational potential) must be a growing function starting from
the brane location. On the brane we assume $\phi (0) = 1$, any
other constant will correspond to an overall re-scaling of the
coordinates. To have convergent transversal volume when $z$ runs
from $0$ to $\infty$ the needed solution $\phi(z)$ of
(\ref{system}) must approach some finite value $a > 1$ at the
infinity.

Boundary conditions are taken in the form
\begin{equation} \label{asimptotics}
\phi (z \rightarrow 0) \approx 1 + z/|c|,~~~~~\phi (z \rightarrow
\infty)\approx a - 1/b |c| z^b ,
\end{equation}
where $b > 0$ is some parameter. This choice corresponds to the
following geometries on the brane and in the transversal infinity:
\begin{eqnarray} \label{geometry}
ds^2(z \rightarrow 0) \approx \eta_{\alpha \beta }(x^\nu)dx^\alpha
dx^\beta + \eta_{ij}dx^i dx^j, \nonumber \\
ds^2 (z \rightarrow \infty)
\approx a^2 \eta_{\alpha \beta }(x^\nu)dx^\alpha dx^\beta +
\frac{1}{z^{b + 1}}\eta_{ij}dx^i dx^j.
\end{eqnarray}

Substitution of the conditions (\ref{asimptotics}) to
(\ref{system}) impose certain relations
\begin{eqnarray} \label{relations}
Aa^5 + Ba + C \simeq 0,~~~~~A + B + C \simeq 0, \nonumber \\
b a^3 - 5Aa^4 - B \simeq 0,~~~~~1 + 5A + B \simeq 0.
\end{eqnarray}
From these relations one can find \cite{GoMi}
\begin{eqnarray} \label{values}
b \approx \frac{4a^3 + 3a^2 + 2a +1}{a^3(a^3 + 2a^2 + 3a + 4)},
\nonumber \\
A = -\frac{\Lambda \epsilon^2c}{40}\approx \frac{1}{a^4 + a^3 +
a^2 + a - 4}, \\
B = \frac{c(\tau + 2L)}{16 M^4} \approx - \frac{a^4+a^3 + a^2 + a
+1}{a^4 + a^3 + a^2 + a - 4}. \nonumber
\end{eqnarray}

Using the relations (\ref{relations}) it can be shown also that
the solution of (\ref{system}) has an inflection point on the
brane $z = 0$ (at the inflection point second derivative of a
function is zero, while the first is not). It means that on the
brane, at the minimum of the transversal gravitational potential
and of the total energy of gravitating system, transversal
curvature $R_{zz}$ is zero. The function $\phi $ has no other
inflection points outside the brane and smoothly grows from $1$ to
its maximal value $a$.

Now it is easy to show that 4-dimensional gravity is localized on
the brane in spite of growing character of transversal potential.
Using formulae for decomposition of the scalar curvature and
determinant
\begin{eqnarray} \label{6R}
^6R = \frac{R}{\phi^2} - 3 \Lambda +
\frac{\tau + 2L}{2\epsilon^2 M^4\phi^4} , \\
\sqrt{^6g} = |c\phi ^{'}|\phi^4\sqrt{-\eta} , \nonumber
\end{eqnarray}
integral of gravitational part of the action (\ref{action}) can be
written in the form:
\begin{eqnarray}  \label{integral1}
S_g = - \int d^6x\sqrt{^6g} \frac{M^4}{2}\left(^6R + 2
\Lambda\right) =  \nonumber \\ = - |c|\epsilon^2\int_{-1}^1dv\int
d^4x\int_0^\infty dz \phi^4
\phi^{'}\sqrt{-\eta}\frac{M^4}{2}\left(\frac{R}{\phi^2} -
\Lambda + \frac{\tau + 2L}{2\epsilon^2 M^4\phi^4}\right) .
\end{eqnarray}

Placing of the minimum of the warp factor at $z=0$ means that
in the frame of the center of the expanding shell-Universe
($x^5 = x^6 = 0$) its walls move towards the transversal
(1+1)-space with a velocity close to the speed of light. In the
considered space (\ref{ansatz}) physical fields are independent
from $v$. Integration of the action over $v$ gives large but finite
universal factor for all kinds of fields (corresponding to the
transversal velocity of the brane) and can be ignored in
calculations. So we must show that physical fields are localized on
the brane only with respect to the coordinate $z$.

Using the relations (\ref{relations}) and
\begin{equation} \label{dz}
\int \sqrt{^6g}dz =
|c|\int_0^\infty \phi^4\phi^{'}\sqrt{-\eta}
= |c|\int_1^a  \phi^4\sqrt{-\eta} d\phi
\end{equation}
one can find that after integration over $z$ last two
terms in (\ref{integral1}) exactly cancel each other. Also we see
that integral over $z$ of remained term is finite in spite of
growing character of $\phi (z) $, since  $\phi (z) $ varies in the
finite range $(1 \div a)$. So 4-dimensional gravity is
localized on the brane and the total action (\ref{action})
reduces to
\begin{eqnarray} \label{integral2}
S = \int d^6x \sqrt{^6g}\left[- \frac{M^4}{2}(^6R + 2 \Lambda) +
^6L\right] \simeq \nonumber \\
\simeq  \int d^4x\int_0^\infty dz \phi^4
\phi^{'}\sqrt{-\eta}\left( -\frac{M^4}{2}\frac{R}{\phi^2} +
\frac{L}{\epsilon^2\phi^4} \right) \simeq \int
d^4x\sqrt{-\eta}\left(-\frac{m^2_P}{2}R + L\right) .
\end{eqnarray}
Appearing in (\ref{integral2}) effective 4-dimensional scale
(Planck's scale)
\begin{equation} \label{m_P}
m^2_P \sim M^4\epsilon^2 a^2
\end{equation}
is constructed from the fundamental scale $M$, the width of our
world $\epsilon$ and the value of the transversal gravitational
potential at the infinity $a$.

For the realistic values (similar to \cite{ADD}) of our physical
parameters
\begin{equation} \label{assumptions}
m^2_P >> M^4\epsilon^2, ~~~~~(\tau + 2L) \sim M^4 > 0,
\end{equation}
from the relations (\ref{values}) follows
\begin{equation} \label{estimations}
a >> 1, ~~~~~ c \sim - 10, ~~~~~ \Lambda > 0, ~~~~~ b \sim 1/a^3,
~~~~~ \epsilon^2 \sim 1/\Lambda a^4.
\end{equation}
Smallness of Newton's constant $\sim 1/m^2_P$ and of the width of
our world $\sim \epsilon $ can be the result of the large values
of the transversal gravitational potential $a$ and of bulk
cosmological constant $\Lambda $.

It must be noted that, since $c$ is negative and $\phi'$ is
Positive, as it is seen from the first equation of (\ref{system})
a suitable solution of our model does not exist in the case of
space-like transversal 2-space of (1+5)-models studied in
\cite{OlVi,ChPo,CoKa,Gr,GhSh,GiMeSh,Gi}.

Now we want to check that in (2+4)-space zero-modes of matter
fields also are localized on the brane with increasing warp
factor. To have self consistent theory we must follow the
assumptions (\ref{extra}) we had used to show localization of
4-dimensional gravity on the brane.

Equation of a massless scalar field in six dimension coupled to
gravity has the form $\partial _A (\sqrt{^6g}g^{AB}\partial_B \Phi
) = 0$. If we take that $\Phi$ is independent from the extra
coordinates we shall receive ordinary 4-dimensional Klein-Gordon
equation and the action of spin $0$ field can be cast to
\begin{equation} \label{action0}
S_\Phi = -\frac{1}{2} \int d^6x \sqrt{^6g}g^{AB}\partial _A \Phi
\partial _B \Phi \simeq -\frac{1}{2} \epsilon^2 \int_1^a
\phi^2d\phi \int d^4x \sqrt{-\eta}\eta^{\mu\nu}\partial _\mu
\Phi\partial _\nu \Phi.
\end{equation}
The localization condition requires the integral over $\phi $ in
(\ref{action0}) to be finite, as it is actually.

The equation and the action of $U(1)$ vector field in the case of
constant extra components $A_i = const $ also reduce to the
4-dimensional Maxwell equations and to the action which is
multiplied by finite integral over extra coordinates
\begin{equation} \label{action1}
S_A = -\frac{1}{4} \int d^6x \sqrt{^6g}g^{AB} g^{MN}F_{AM}F_{BN}
\simeq -\frac{1}{4} \epsilon^2 \int_1^a d\phi \int d^4x
\sqrt{-\eta}\eta^{\mu\nu}\eta^{\alpha\beta}F_{\mu\alpha}F_{\nu\beta}.
\end{equation}

From the convergent character of the volume element (\ref{dz}) and 
formulae (\ref{action0}) and (\ref{action1}) it easy to see that 
localization of the Abelian-Higgs model (investigated in the papers 
\cite{GiMeSh,Gi} for the signature $(1+5)$) is the particular example 
of our model. In addition here we have localization of zero modes of 
spinor fields also.

In the case of spinor fields we shall introduce the vierbien
$h^{\bar{M}}_M$, where $\bar{M}, \bar{N}, ...$ denote local
Lorentz indices. Relation between the curved gamma matrices
$\Gamma^M$ and the flat gamma ones $\gamma^{\bar{M}}$ is given by
the formula $\Gamma^M = h^M_{\bar{M}}\gamma^{\bar{M}}$, so
\begin{equation} \label{gamma}
\Gamma_\mu = \phi \gamma_\mu , ~~~~~ \Gamma_i = \sqrt{|c|\phi
'}\gamma_i .
\end{equation}
The spin connection is defined as
\begin{equation} \label{spin1}
\omega^{\bar{M}\bar{N}}_M = \frac{1}{2} h^{N\bar{M}} (\partial_M
h^{\bar{N}}_N - \partial_N h^{\bar{N}}_M) - \frac{1}{2}
h^{N\bar{N}}(\partial_M h^{\bar{M}}_N - \partial_N h^{\bar{M}}_M)
- \frac{1}{2} h^{P\bar{M}}h^{Q\bar{N}}(\partial_P h_{Q\bar{R}} -
\partial_Q h_{P\bar{R}})h^{\bar{R}}_M .
\end{equation}
The non-vanishing components of the spin-connection for the
background metric (\ref{ansatz}) are
\begin{equation} \label{spin2}
\omega^{\bar{M}\bar{N}}_\nu = (\delta^{i\bar{N}}
\delta^{\bar{M}}_\nu - \delta^{i\bar{M}}\delta^{\bar{N}}_\nu
)\partial_i\phi /\sqrt{|c|\phi '} , ~~~~~
\omega^{\bar{M}\bar{N}}_j =
(\delta^{i\bar{N}}\delta^{\bar{M}}_j -
\delta^{i\bar{M}}\delta^{\bar{N}}_j )\partial_i\sqrt{\phi
'}/\sqrt{\phi '}.
\end{equation}

Therefore, the covariant derivatives have the form
\begin{equation} \label{covariant}
D_\mu\Psi = ( \partial_\mu +  \Gamma^j \Gamma_\mu \partial_j\phi
/2\phi) \Psi , ~~~~~ D_i\Psi = (\partial_i +  \Gamma^j \Gamma_i
\partial_j\sqrt{\phi '}/2\sqrt{\phi '}) \Psi .
\end{equation}

We are looking for the solution in the form $\Psi (x^A) = \psi
(x^\nu)H(x^j)$, where $\psi $ satisfies the massless 4-dimensional
Dirac equation $\gamma^\nu \partial_\nu \psi = 0$ . Then
6-dimensional Dirac equation reduces to
\begin{equation} \label{equationH}
\left( \partial_j - 2 \partial_j \phi / \phi - \partial_j
\sqrt{\phi '} / 2 \sqrt{\phi '} \right) H(x^j) = 0
\end{equation}
The solution of this equation with unit integration constant is
\begin{equation} \label{H}
H(x^j) = \phi^2 (\phi ')^{1/4} .
\end{equation}
Then the action of spin $1/2$ field takes the form
\begin{equation} \label{action1/2}
S_\Psi = \int d^6 x \sqrt{^6g}\bar{\Psi} i \Gamma^A D_A \Psi
\simeq \epsilon^2 \int_0^\infty dz \phi^7 (\phi ')^{3/2}\int d^4 x
\sqrt{-\eta } \bar{\psi} i \gamma^\nu
\partial_\nu \psi  .
\end{equation}
Since $\phi $ and $\phi '$ are monotone and finite functions of
$z$ the integral over $z$ in (\ref{action1/2}) is finite. So
massless Dirac fermions are also localized on the brane.

When we consider interaction of scalars, or fermions with the
electromagnetic field we must make usual replacements
\begin{equation} \label{interaction}
\partial_i \rightarrow \partial_i - iA_i , ~~~~~\Psi \rightarrow
e^{iA_i x^i} \Psi
\end{equation}
in above formulae for localization. Here $x^i$ are coordinates of
transversal (1+1)-space and $A_i$ are constant extra components of
electromagnetic field. 

To summarize, in this paper it is shown that for the realistic
values of the fundamental scale and the brane stress-energy, there
exists a non-singular static solution of (2+4)-dimensional
Einstein equations. This solution provides gravitational trapping
of the 4-dimensional gravity and the matter on the brane without
extra $\delta$-like sources. In contrast to Randall-Sundrum's
case, the factor responsible for this trapping is the growing
away from the brane gravitational potential, but has a convergent
volume integral, although the transversal 2-space is infinite.
Study of the fluctuation of the metric, which is crucial for the
stability of the model, will be the subject of future
investigations.

\end{document}